# Intrinsic radiation from Poynting jets and magnetized collisionless shocks: implications for prompt GRB emissions

E. P. LIANG and K. NOGUCHI
*Rice University, Houston, TX 77005-1892*

**Summary**

We summarize latest PIC simulation results on the radiation from Poynting jets and strongly magnetized collisionless shocks when a Poynting jet runs into cold ambient medium. We find that in all cases the radiative power output is much below that predicted by synchrotron radiation and the critical frequency is also much lower than the synchrotron critical frequency. We discuss the implications for the interpretation of GRB prompt emission data.

## 1. Introduction

Two popular paradigms of GRB energy source are relativistic hydrodynamic (HD) outflows [1], and electromagnetic (EM) or Poynting flux (PF)-dominated outflows [2], driven by the formation of a black hole or magnetar in a collapsar or compact-objects merger. In either paradigm the challenge is to find robust mechanisms which efficiently convert HD or EM energy into the relativistic kinetic energy of nonthermal particles and radiation. A popular model of the HD paradigm is the synchrotron shock model, where unsteady flow leads to dissipation via internal shocks in the prompt phase and snowplowing of the CSM/ISM leads to external shock emission in the afterglow phase. In collisionless shocks (CS), Weibel instability [3] has been invoked to generate turbulent magnetic fields, leading to diffusive electron acceleration followed by "jitter" radiation [4], though inverse Comptonization in dense photon environments [4] has also been suggested. In the PF paradigm, particle acceleration and radiation are driven directly by large-scale EM fields via collective plasma processes [2]. A hybrid scenario is EM-dominated shocks, in which the internal/external shocks are mediated by strong ordered EM fields.

In all scenarios, particle acceleration and radiation ultimately come down to plasma kinetics, which are best modeled using large-scale Particle-in-Cell (PIC) codes [5]. In the past few years, using the most advanced PIC codes, we have performed hundreds of simulations of PF and CS scenarios [6,8]. Here we summarize the latest results on the radiation output of sample PF and CS models. We compute the instantaneous radiation power output of each particle directly from the particle and field data of the PIC simulations using $P_{rad} = 2e^2(F_{\parallel}^2 + \gamma^2 F_{+}^2)/3c$ where $F_{\parallel}$ is the force parallel to velocity **v** and $F_{+}$ is the force orthogonal to **v** [4], and compare them to GRB data. The most significant finding is that *in both PF and CS models the radiation from the most energetic particles are far from classical synchrotron radiation, and the typical power output is many orders of magnitudes below that of synchrotron radiation for a given magnetic field and Lorentz factor*. In the PF case, the critical frequency is also much lower than the synchrotron critical frequency. These results have far reaching implications for the interpretation of GRB output and spectra [7].

## 2 Radiation from PF Models

The physics of direct-drive PF acceleration have been discussed in previous publications [6, 8] and will not be repeated here. (Several powerpoint presentations and movies are available at our websites: http://spacibm.rice.edu/~liang/picsim and http://spacibm.rice.edu/~knoguchi). Direct PF acceleration of plasmas fall into two main categories: front-loaded or back-loaded. In the former case, which we will call leading Ponderomotive accelerator (LPA), the EM ponderomotive force snowplows the electrons ahead of it without penetrating beyond the skin depth, as in laboratory laser-target interactions. In this case the PF momentum is shared by all upstream particles, so the Lorentz factor of the particles is limited by max($\Omega_e/\omega_{pe}$, $a_o^2/2$) where $\Omega_e$ is the electron gyrofrequency, $\omega_{pe}$ is the electron plasma frequency and $a_o = eA/mc$ is the dimensionless vector potential. In the latter case, which we call trailing Ponderomotive accelerator (TPA, this term replaces the acronym DRPA used in our early publications) the plasma-loaded EM pulse pulls the trailing plasma behind it via self-induced J x B force. The PF continuously accelerate the fastest particles but gradually leave slower particles behind. Decreased plasma loading allows both the PF and residual fast particles to accelerate over time, reaching Lorentz factors far exceeding $\Omega_e/\omega_{pe}$ or $a_o^2/2$ in our PIC simulations. Ultimately the Lorentz factor is limited by radiation damping or dephasing (self-induced or externally induced). We have studied radiation by both TPA and LPA. For a given input PF amplitude, TPA particles radiates much more broadly than LPA since all particles are imbedded in the strong fields. Even for TPA, $P_{rad}$ is many orders of magnitude below that of classical synchrotron radiation with the same B and $\gamma$. This is because the fast TPA particles see a comoving B field. Both the net Lorentz force and particle velocity make only small angles with the Poynting vector. We find that radiation from the highest energy TPA particles can be approximated to first order by the analytic formula $P_{analytic} \sim r_o^2 c\ B^2 p_+^2 \sin^2\alpha$, where $r_o$ is classical electron radius, $p_+$ is total momentum *orthogonal* to the Poynting vector **k** (including component of **p** ∥ to **B**!), and $\alpha$ is the angle between **v** and **k**. Fig.1 compares the numerical $P_{rad}$ versus $P_{analytic}$, showing excellent correlation (the scatter is due to initial momentum terms not included in the above formula). Intuitively, high $\gamma$ particles are moving mainly along **k**, so $p_+ \ll \gamma$ and $\sin\alpha \ll 1$. Hence $P_{rad}$ is $\ll$ classical synchrotron power $P_{syn} \sim r_o^2 c\ B^2 \gamma^2$. Similarly, the comoving B causes only small curvature in the particle tracks. So the critical radiation frequency which is related to the duration of sweep of the photon beam across the observer at infinity [4] can be shown to approximate $\omega_{cr} \sim \Omega_e\ \gamma^2 \sin^2\alpha \ll$ the synchrotron critical frequency $\omega_{cr} \sim \Omega_e\ \gamma^2$.

In all PF models, the late-time $P_{rad}$ asymptotes to a constant value which depends on the initial ejecta temperature and magnetic field. The hotter the initial ejecta, the higher the asymptotic $P_{rad}$ (Fig.2). We find that the asymptotic $P_{rad}$ scales roughly as $p_o^2$ where $p_o$ is the initial average particle momentum. We also find that the asymptotic $P_{rad}$ scales with initial PF field $B_o$ roughly as $B_o^n$ with n~2-3 (Fig.3). In most PF runs the asymptotic particle spectrum has a power law index near 3.5 (Fig4). This is consistent with the observed gamma-ray spectrum of most high energy sources in the optically-thin non-cooling limit [4].

### 3. Radiation from CS models

(1). In CS models, $P_{rad}$ is strongly dependent on B at the shock. For transverse EM-dominated shocks (Alfven speed $v_A > c$), we obtain values similar to the PF case for ejecta electrons, but factors of 10 lower for shocked ambient electrons (Fig.5). For weakly magnetized shocks, simulations suggest that similar scaling exists but we have not yet derived an analytic formula. In nonmagnetic shocks we find that the highest energy particle are mainly accelerated by plasma wave electric fields, not Fermi scattering as had been postulated. In this case $P_{rad}$ is much lower than even the PF formula, when B refers to the downstream turbulent field generated by Weibel instability [3]. This is because the high γ particles are mostly accelerated away from the Weibel-generated field region.

2). In CS models of e+e- shocking e-ion plasmas (Fig.6), we find that the shocked ambient electrons radiate roughly at the same level as in e+e- shocking e+e- plasmas for a given B field of the ejecta.

3). In all magnetized PF and CS cases, the radiation output is strongly linearly polarized as expected [8,9].

## 4. Implications for prompt GRB emissions

Current models of GRB emissions, whether they are driven by HD or PF, assume that radiation occurs in-situ with the particle acceleration. As we show above, the in-situ radiation in all our PIC simulations are many orders of magnitude below that of synchrotron radiation, for a given local magnetic field and particle Lorentz factor. Physically, this result is not surprising, since the condition is most favorable to achieve the highest energy when the in-situ Lorentz force is most aligned with the particle momentum, whereas synchrotron radiation occurs when the particle momentum is mostly transverse to the Lorentz force. Also the particle momentum distributions in our PIC simulations are highly anisotropic, with the largest component along the Poynting vector in the PF case and normal to the shock front in the CS case. This suggests that particles will lose much less energy per unit distance traveled and achieve much higher Lorentz factors than conventional synchrotron models. Unless these ultrarelativistic particles eventually finds a "beam dump" to convert their energy into synchrotron (if the beam dump is a strongly magnetized plasma) or inverse Compton radiation (if the beam dump is a dense soft photon field), the accelerator will traverse large distances before the HD or PF energy is fully radiated away. On one hand this solves the rapid cooling problem of synchrotron shock models. On the other hand we need to completely revise the emission region parameters which are usually based on the synchrotron model. The results of our PIC simulations suggest that we need to completely revise the "standard" scenario based on alternative emission mechanisms much less efficient than synchrotron radiation. These and other implications for the interpretation of GRB prompt emissions, include the Amati-Ghirlanda relations [10, 11] will be addressed in future papers.

**Figure captions**

**Fig.1** Scatter plot of radiative power $P_{rad}$ radiated by each particle in a Poynting jet expanding from grid center compared with the analytic formula $P_{analytic} \sim r_o^2 c\ B^2 \gamma^2 \sin^4\alpha$, shows good correlation, especially for high $\gamma$ particles. We plot only one out of every 8 particles in the simulations in all figures. Units are arbitrary but normalized to a standard value for all runs.

**Fig2** Evolution of $P_{rad}(x,t)$ vs. distance for Poynting jets of different initial $kT_o$ shows that higher initial temperature (a=5 MeV, b= 5 keV) leads to higher radiative power per particle and broader emission regions.

**Fig.3** Evolution of $P_{rad}(x,t)$ vs. distance for Poynting jets of different initial magnetic fields $v_A/c$ (a=$10^2$, b=$10^3$, c=$10^4$) but fixed initial temperature (kT = 5 keV) shows that $P_{rad}(x,t)$ scales as $B^n$ where n~2-3. In all cases $P_{rad}$ reaches ~constant level at late times, despite the continuous growth in particle Lorentz factors.

**Fig.4** Particle energy spectra from different Poynting jets show robust power-law index ~ 3.5. This particle index of ~ 3.5 is consistent with the high-energy photon power-law index of most GRBs, AGNs and pulsars [7].

**Fig.5** (a) Details of magnetized shock structure when a Poynting jet runs into cold e+e- ambient plasma. Note that the ejecta runs upstream of the shocked ambient electrons, unlike MHD shocks, and the Poynting jet magnetic field is strongly suppressed inside the transition region. (b) Evolution of $P_{rad}(x,t)$ vs. distance for the shocked ambient e-. Note that the initial shock produces high radiative power, followed by a rapid decrease before recovering to an almost constant $P_{rad}$ level at late times.

**Fig.6** (a) Details of magnetized shock structure when a Poynting jet runs into cold e-ion plasma. The transition region is much thicker than that of Fig.4 due to charge separation between shock electrons and ions. (b) Evolution of $P_{rad}(x,t)$ vs. distance for ejecta electrons and for shocked ambient electrons (c). Note that the radiating zone is now much broader than in Fig.4, due to charge separation between shocked e- and ions. However the magnitude of $P_{rad}$ is comparable to that of Fig.4. The shock is most radiative at the beginning, followed by rapid decay, before it recovers to a constant asymptotic level.

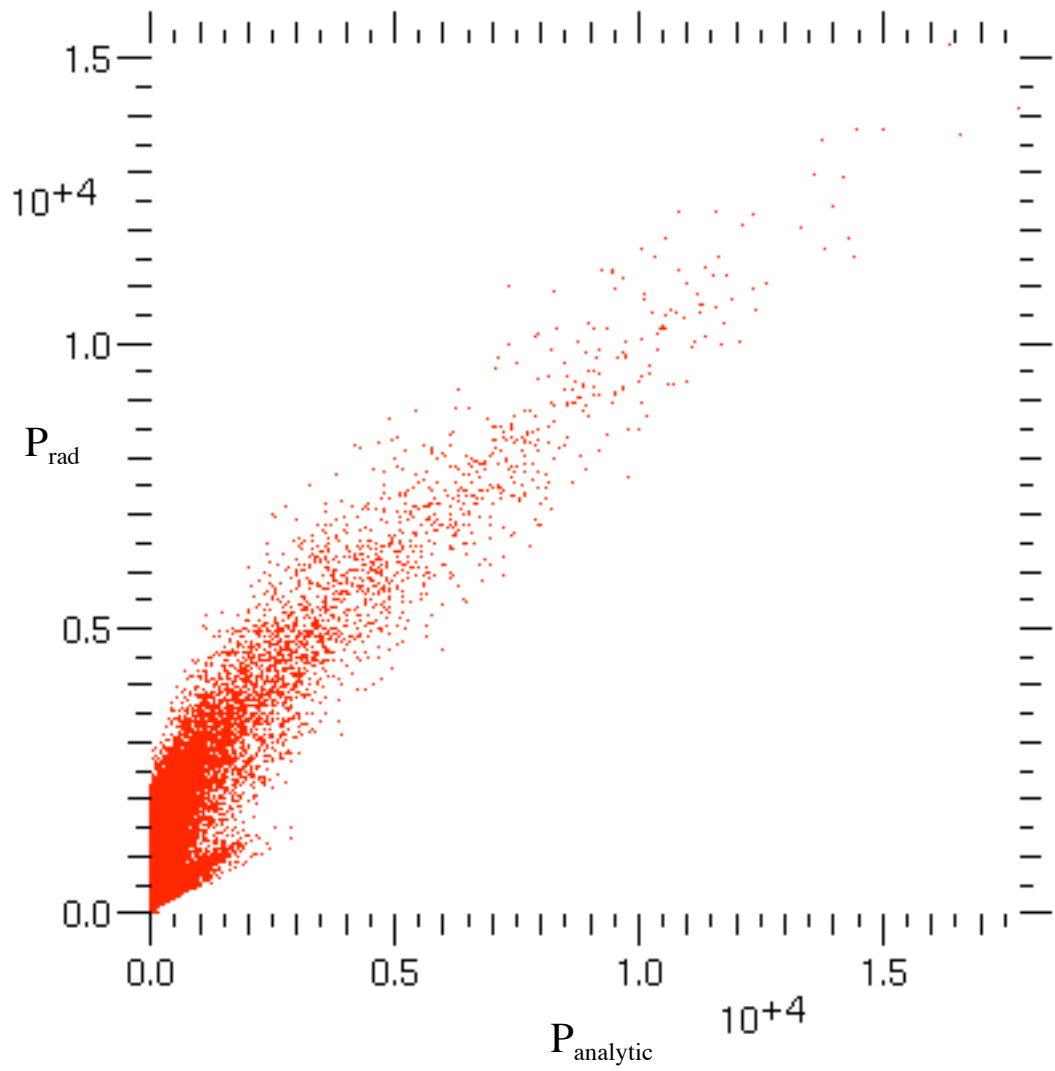

Fig1

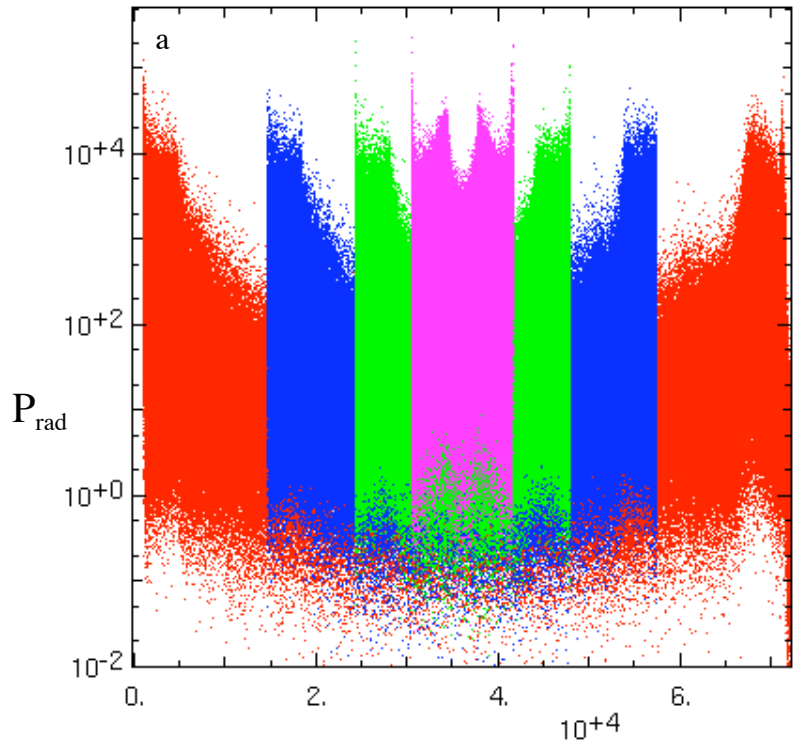
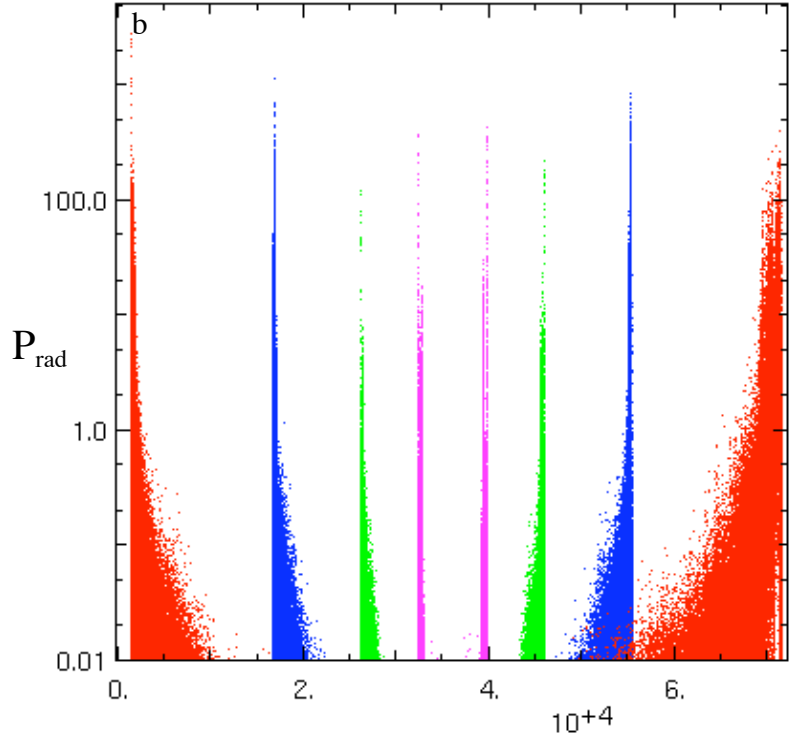

Fig2

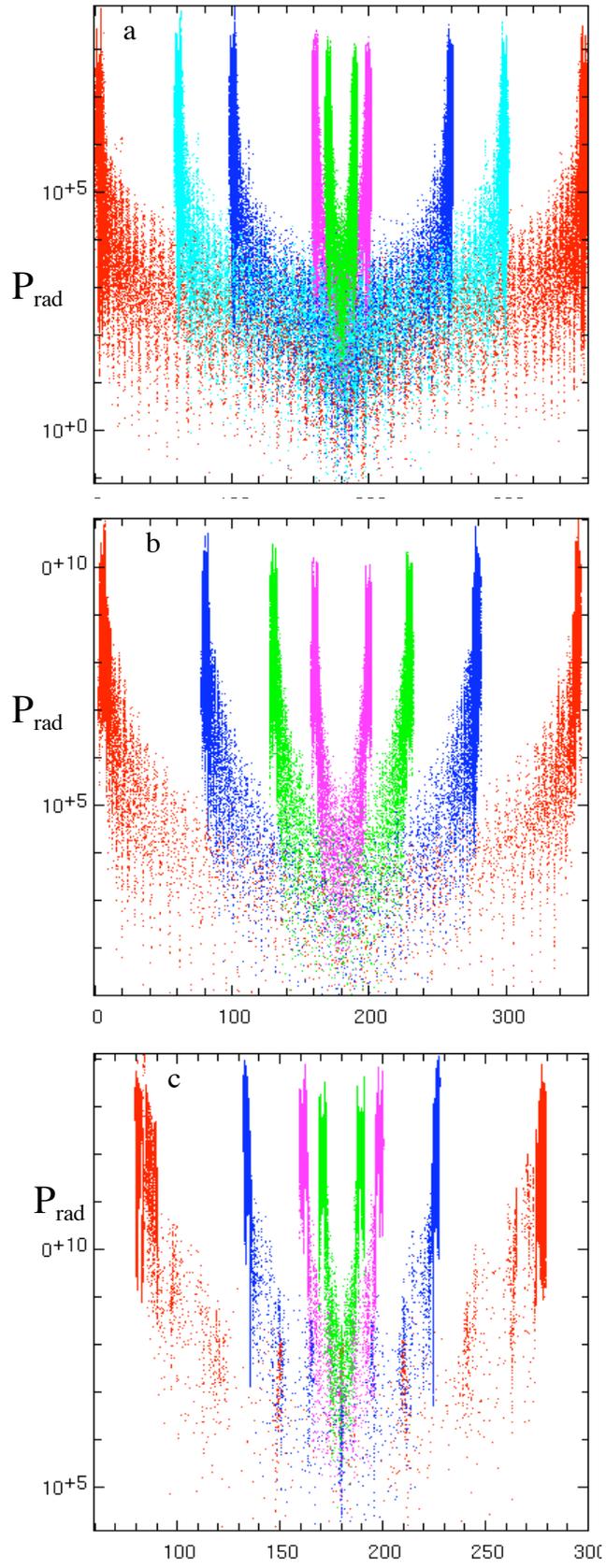

Fig.3

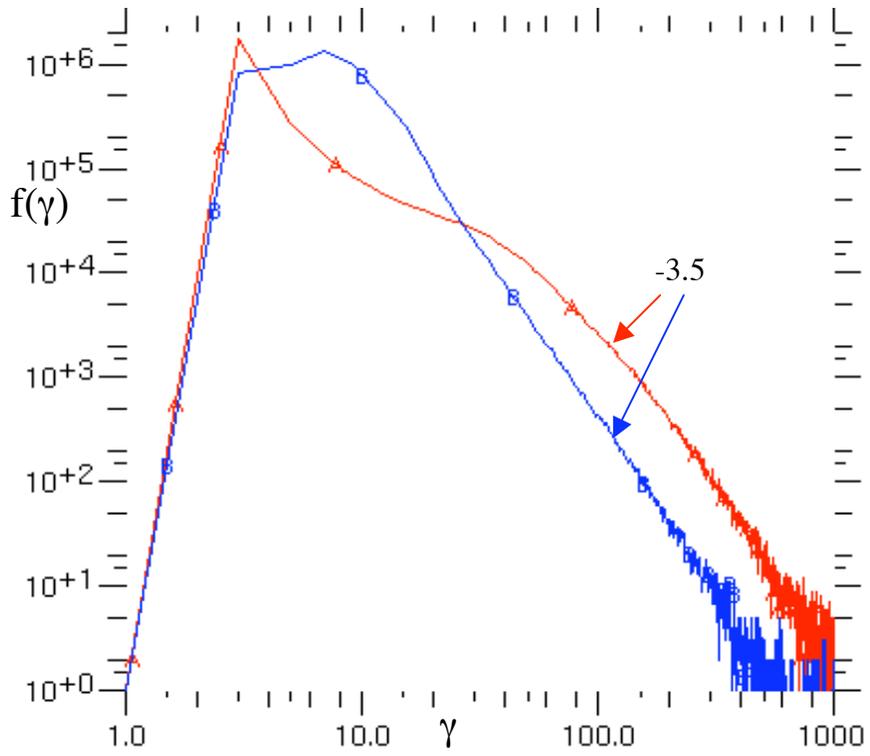

Fig.4

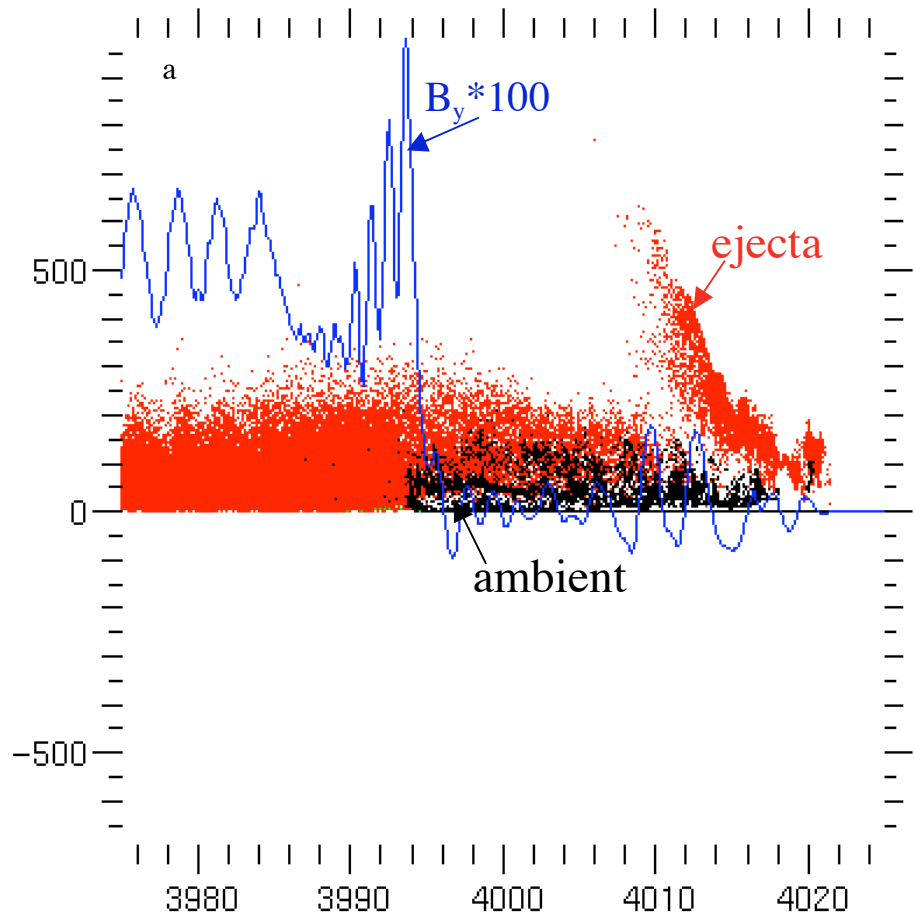
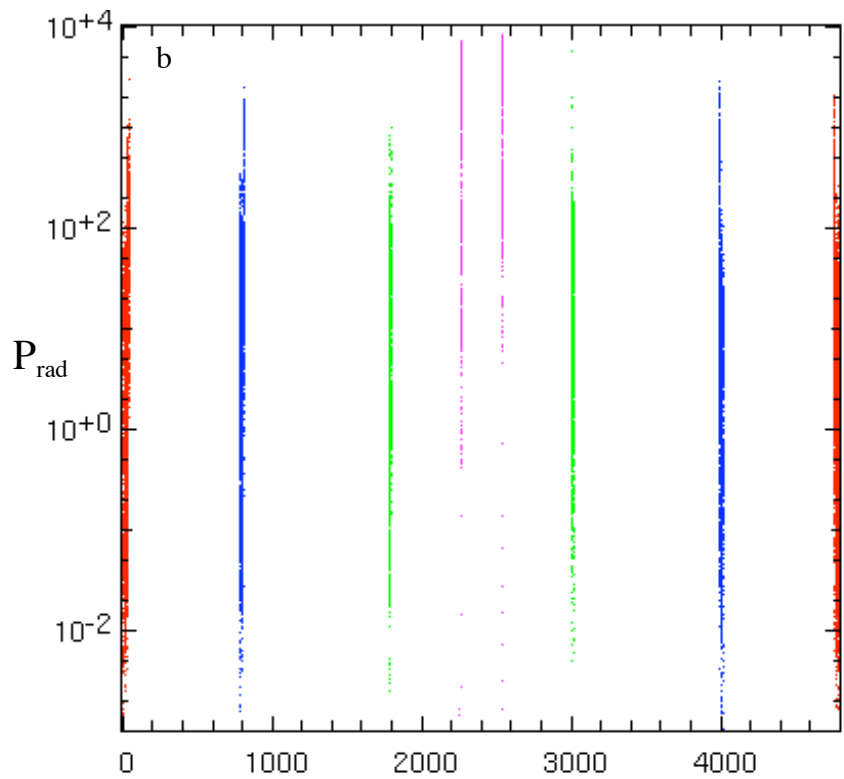

Fig.5

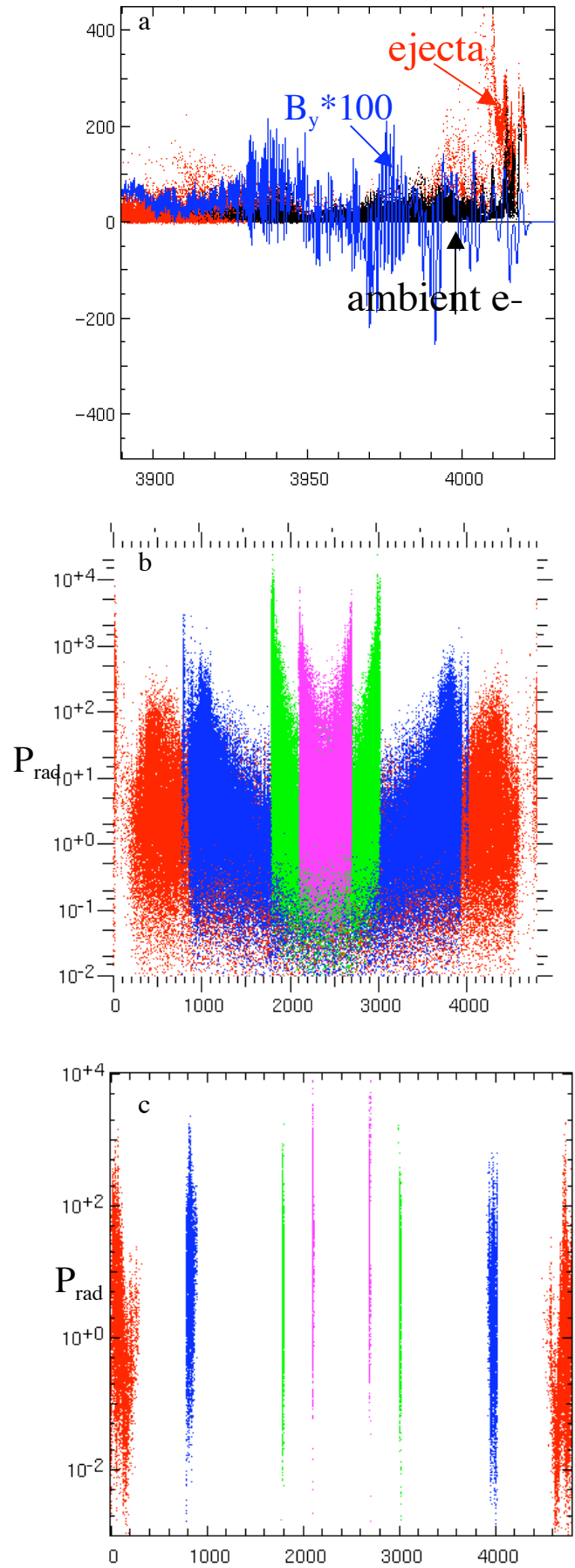

Fig.6